\documentclass{ws-ijmpa}
\usepackage{graphicx}
\usepackage{epstopdf}
\usepackage{multirow}

\begin{document}

\markboth{Ziwen Fu}
{Studying $\kappa$ meson with a MILC fine lattice}

\title{Studying $\kappa$ meson with a MILC fine lattice}

\author{Ziwen Fu}

\address{
Key Laboratory of Radiation Physics $\&$ Technology {\rm (Sichuan University)},
Ministry of Education;
Institute of Nuclear Science $\&$ Technology, Sichuan University,
Chengdu 610064, P. R. China.\\
fuziwen@scu.edu.cn
}
\maketitle

\begin{abstract}
Using the lattice simulations in the Asqtad-improved staggered fermion formulation
we compute the point-to-point $\kappa$ correlators, which are analyzed
by the rooted staggered chiral perturbation theory (rS$\chi$PT).
After chiral extrapolation, we secure the physical $\kappa$ mass with $835\pm93$~MeV,
which is in agreement with the BES experimental results.
The computations are performed using a MILC $2+1$ flavor
fine gauge configuration at a lattice spacing of $a \approx 0.09$~fm.

\end{abstract}
\ccode{PACS numbers: 12.38.Gc,  11.15.Ha}

\section{Introduction}
\label{sec_intro}
In 2012, the Particle Data Group (PDG)\cite{Beringer:1900zz}
lists the $K_0^*(800)$ meson $I(J^P)=\frac{1}{2}(0^+)$,
which is normally called $\kappa$ meson, with a mass of $682\pm29$~MeV.
Many experimental analyses\cite{Bugg:2005xx,Bugg:2009uk,Aitala:2002kr,Ablikim:2010ab,Ablikim:2010kd,Ablikim:2005ni,Alde:1997ri,Xiao:2000kx,Cawlfield:2006hm,Bonvicini:2008jw}
strongly support its subsistence,  and
the recent BESII analyses gives its mass about $826 \pm 49$~MeV.\cite{Ablikim:2010ab}
Nonetheless, the existence of $\kappa$ meson is still slightly  debatable.\cite{Beringer:1900zz}

It is not decided whether $\kappa$ is traditional $\bar qq$ or tetraquarks
$\bar q\bar qqq$.\cite{Prelovsek:2010kg,Prelovsek:2008rf,Alford:2000mm,Loan:2008sd,Wagner:2013nta,Alexandrou:2012rm}
The tetraquarks interpretations of the scalar mesons
are able to realize
the experimental mass ordering $m_{a_0(980)}>m_{\kappa}$ 
since the $I=1$ state ($\bar u\bar s sd$) is heavier than
the $I=1/2$ state ($\bar u\bar d ds$) due to $m_s>m_d$,
whereas the conventional $\bar ud$ and $\bar us$ states
can with difficulty interpret the observed mass ordering.
Sasa Prelovsek {\it et al.} found that $\kappa$ meson have large
tetraquark component,\cite{Prelovsek:2010kg,Prelovsek:2008rf}
whereas M. Wagner {\it et al.} demonstrated that $\kappa$ meson
does not have sizeable tetraquark component,\cite{Wagner:2013nta}
and they even plan to combine four quarks with traditional
quark-antiquark operators.\cite{Alexandrou:2012rm}
Therefore, the lattice studies have definitely not yet reached consensus
whether the $\kappa$ meson is
tetraquark or conventional $\bar{q}q$ meson.
This issue can be partially solved if the mass of scalar $\bar{q}q$
state with $I=1/2$ can be robustly calculated on the lattice.
We refer to this state as $\kappa$ meson in this paper.

To date, only a couple of lattice studies on $\kappa$ mass
(to be specific, we here mean the $u\bar{s}$ scalar meson)
have been published.
Prelovsek {\it et al.} delivered a rough calculation of
the $\kappa$ mass as $1.6$~GeV through extrapolating
the $a_0$ mass.\cite{Prelovsek:2004jp}
In the quenched approximation,
Mathur {\it et al.}\cite{Mathur:2006bs} examined
the scalar $u\bar{s}$ meson and estimated the $\kappa$ mass
to be $1.41$~GeV with taking off the $\pi\eta^{\prime}$ ghost.
The UKQCD Collaboration\cite{McNeile:2006nv}
indicated $\kappa$ mass about $1.0-1.2$~GeV
using the dynamical $N_{f}=2$ sea quarks.
The full QCD simulations on $\kappa$ meson
are carried out by SCALAR group\cite{Kunihiro:2003yj,Kunihiro:2004ga}
using strange quark as  a valence approximation,
which indicated that $\kappa$ mass is around $1.8$~GeV.
A quenched QCD computation was conducted\cite{Wada:2007cp}
using the Wilson fermions,
and $\kappa$ mass is estimated to be around $1.7$~GeV.

With the $2+1$ flavors of the Asqtad-improved staggered sea quarks,
we handled the $u$ quark as a valence approximation quark,
whereas the valence strange quark mass is set to be the physical
mass,\cite{Bazavov:2009bb,Bernard:2010fr} and secured $\kappa$ mass with
$826 \pm 119$~MeV,\cite{Fu:2011zz}
unfortunately, we neglected the taste-symmetry breaking
due to the staggered scheme.\cite{Prelovsek:2005rf}
We addressed this issue by extending the analyses
of the scalar $f_0$ and  $a_0$
mesons\cite{Fu:2011gw,Bernard:2007qf,Fu:2011zzh,Fu:2011zzl}
to scalar $\kappa$ meson, and calculated it on a MILC coarse ($a=0.12$ fm) lattice ensemble.
After chirally extrapolating $\kappa$ mass to physical point,
we gained the $\kappa$ mass $828\pm97$~MeV (Ref.~\refcite{Fu:2011xb})
with the consideration of bubble contribution.\cite{Prelovsek:2005rf}
Since the occurrence of bubble contribution is
an aftermath of the fermion determinant,
the bubble term in the staggered chiral perturbation theory (S$\chi$PT)
furnishes an instrumental interpretation of the lattice artifacts
caused by the fourth-root procedure.\cite{Fu:2011gw,Bernard:2007qf}
We realize that the bubble contribution should be considered
in the spectral process of the $\kappa$ correlator
for the MILC medium coarse ($a\approx0.15$~fm)
and coarse ($a\approx0.12$~fm) lattice ensembles.

Moreover, the rS$\chi$PT forecasts  additionally that these lattice artifacts
disappear in the continuum limit,
just remaining the physical thresholds.\cite{Fu:2011xb}
To verify this prediction, we here conduct a quantitative
comparison of the calculated $\kappa$ correlators with the prognostications of rS$\chi$PT
at a MILC fine ($a\approx0.09$~fm) lattice ensemble.
As we expected, the lattice artifacts are indeed further suppressed,
and our simulation result for the $\kappa$ masses
illustrated that the bubble contribution is negligible
in the accuracy of statistics for our chosen MILC fine lattice ensemble.

\section{Pseudoscalar meson taste multiplets}
\label{sec_taste_multiplets}
In Refs.~\refcite{Fu:2011gw,Bernard:2007qf}
we reviewed the rooted staggered chiral perturbation theory
using the replicated trick.\cite{Aubin:2003rg}
The tree-level pseudoscalar meson masses
are given by\cite{Bernard:2007qf,Aubin:2003mg}
\begin{equation}
M_{x,y,b}^2 = \mu(m_x + m_y ) + a^2\Delta_b,
\label{tree_splitting}
\end{equation}
where $\mu$ is a low-energy chiral coupling constant,
$b$ is taste index,
and the $a^2\Delta_b$ originates from the taste-symmetry breaking.
The $m_x$ and $m_y$ are two valence quark masses of pseudoscalar meson.
Since we study with the degenerate $u$ and $d$ quarks,
it is customary to make use of the shorthand notations
\begin{eqnarray}
M^2_{Ub} &\equiv& M_{\pi_b} = 2\mu m_x + a^2 \Delta_b, \cr
M^2_{Sb} &\equiv& M_{ss,b}  = 2\mu m_s + a^2 \Delta_b, \\
M^2_{Kb} &\equiv& M_{K_b}   = \mu (m_x + m_s) + a^2 \Delta_b, \nonumber
\end{eqnarray}
where $M_U$ is the Nambu-Goldstone pion mass,
$M_K$ is the Nambu-Goldstone kaon mass, and
$M_S$ is the mass of an imaginary flavor nonsinglet
meson $s\bar{s}$.\cite{Aubin:2004wf}

Given a large anomaly parameter $m_0$, we have\cite{Aubin:2004wf}
\begin{equation}
M_{\eta, I}^2 = \frac{1}{3}M_{UI}^2 + \frac{2}{3}M_{SI}^2, \hspace{0.5cm}
M_{\eta^\prime, I} =  {\cal O}(m_0^2),
\end{equation}
and in the taste-axial-vector sector
\begin{eqnarray}
M^2_{\eta A}        &=& \frac{1}{2}[M^2_{UA}+M^2_{SA}+\frac{3}{4}\delta_A-Z_A], \cr
M^2_{\eta^\prime A} &=& \frac{1}{2}[M^2_{UA}+M^2_{SA}+\frac{3}{4}\delta_A+Z_A], \\
Z^2_A               &=& (M^2_{SA}-M^2_{UA})^2-\frac{\delta_A}{2} (M^2_{SA}-M^2_{UA})+\frac{9}{16}\delta^2_A, \nonumber
\label{eq_etaAV}
\end{eqnarray}
and similarly for $V \to A$,
where $\delta_V$ is a hairpin coupling of 
two taste-vector mesons.\cite{Aubin:2004fs}
In the taste-pseudoscalar and taste-tensor sectors,
the $\eta_b$ and $\eta'_b$ masses are given by
\begin{equation}
M_{\eta, b}^2 = M_{Ub}^2 ; \qquad  M_{\eta', b}^2 = M_{Sb}^2 .
\end{equation}
In Table~\ref{tab_L0}, we tabulate the masses of the calculated
taste multiplets  using the values of $\delta_A$ and $\delta_V$ 
determined by MILC Collaboration.\cite{Aubin:2004wf,Aubin:2004fs}
\begin{table}[h!]
\tbl{ \label{tab_L0}
The masses of the pseudoscalar meson in lattice units for the MILC fine
($a = 0.09$ fm) lattice ensemble
with $\beta = 7.09$, $am_{ud}^{\prime} = 0.0062$, $am_s^{\prime} = 0.031$.
}
{ \begin{tabular}{@{}cccccc@{}}  
\toprule
$am_x$ & taste($B$) & $a\pi_B$ & $aK_B$ & $a\eta_B$ & $a\eta^\prime_B$ \\
\colrule
\multirow{5}*{$0.0062$}
&P & $0.1480$ & $0.2339$ & $0.1480$ & $0.2951$ \\
&A & $0.1645$ & $0.2447$ & $0.1529$ & $0.3009$ \\
&T & $0.1742$ & $0.2513$ & $0.1742$ & $0.3091$ \\
&V & $0.1818$ & $0.2566$ & $0.1778$ & $0.3123$ \\
&I & $0.1922$ & $0.2642$ & $0.2836$ & $\cdots$ \\
\hline
\multirow{5}*{$0.0093$}
&P & $0.1807$ & $0.2448$  &$0.1807$ & $0.2951$ \\
&A & $0.1944$ & $0.2551$  &$0.1847$ & $0.3009$ \\
&T & $0.2027$ & $0.2615$  &$0.2027$ & $0.3091$ \\
&V & $0.2093$ & $0.2666$  &$0.2058$ & $0.3223$ \\
&I & $0.2185$ & $0.2739$  &$0.2898$ & $\cdots$ \\
\hline
\multirow{5}*{$0.0124$}
&P & $0.2080$ & $0.2554$  & $0.2080$ & $0.2951$ \\
&A & $0.2200$ & $0.2653$  & $0.2114$ & $0.3010$ \\
&T & $0.2274$ & $0.2714$  & $0.2274$ & $0.3091$ \\
&V & $0.2332$ & $0.2763$  & $0.2301$ & $0.3123$ \\
&I & $0.2415$ & $0.2833$  & $0.2959$ & $\cdots$ \\
\hline
\multirow{5}*{$0.0155$}
&P & $0.2320$ & $0.2655$  & $0.2320$ & $0.2951$ \\
&A & $0.2429$ & $0.2750$  & $0.2350$ & $0.3010$ \\
&T & $0.2495$ & $0.2809$  & $0.2495$ & $0.3091$ \\
&V & $0.2549$ & $0.2857$  & $0.2520$ & $0.3123$ \\
&I & $0.2625$ & $0.2925$  & $0.3018$ & $\cdots$ \\
\hline
\multirow{5}*{$0.0186$}
&P & $0.2539$ & $0.2754$  & $0.2539$ & $0.2951$ \\
&A & $0.2639$ & $0.2846$  & $0.2565$ & $0.3012$ \\
&T & $0.2700$ & $0.2903$  & $0.2700$ & $0.3091$ \\
&V & $0.2750$ & $0.2949$  & $0.2723$ & $0.3123$ \\
&I & $0.2820$ & $0.3015$  & $0.3076$ & $\cdots$ \\
\botrule
\end{tabular} }
\end{table}

\section{The $\kappa$ correlator from S$\chi$PT}
\label{sec:correlators}
In term of the terminology of
replica recipe\cite{Aubin:2003mg,Damgaard:2000gh} and
by matching the scalar $\kappa$ correlator
in the low-energy chiral effective theory with the staggered fermion QCD,
we derived the bubble contribution to the $\kappa$ meson.\cite{Fu:2011xb}
Here we review some results needed for the present study.

In order to acquire the proper number of the quark species,
we execute the fourth-root procedure, and utilize an interpolation operator
with $I(J^{P})=\frac{1}{2}(0^{+})$ at source and sink,
\begin{eqnarray}
{\cal O}(x)  \equiv
\frac{1}{\sqrt{n_r}} \sum_{a, g}\bar s^a_g( x ) u^a_g(x) ,
\end{eqnarray}
where $a$, $g$ and $n_r$ are color index, taste replica index,
      and the number of the taste replicas, respectively.
The time slice $\kappa$ correlator $C(t)$ can be measured by
\begin{eqnarray}
C(t) &=&
\frac{1}{n_r}
\sum_{ {\mathbf{x}}, a, b } \sum_{g, g'}
\left\langle \bar s^{b}_{g'}({\mathbf{x}}, t)
u^{b}_{g'}({\mathbf{x}}, t) \bar u^{a}_{g }({\bf 0}, 0) s^{a}_{g }({\bf 0}, 0) \right\rangle , \nonumber
\label{EQ.kappa}
\end{eqnarray}
where ${\bf 0}$ and ${\mathbf{x} }$ are spatial points of
the $\kappa$ state at source and sink, respectively.
After conducting Wick contractions of fermion fields,
and carrying out the summation over the taste index,~\cite{Fu:2011zz}
we arrive at
\begin{equation}
\label{CCEQ_kappa}
C(t) = \sum_{ {\mathbf{x} } } (-1)^x
\left\langle \mbox{Tr}
[M^{-1}_u( {\mathbf{x} },t;0,0) M^{-1^\dag}_s({\mathbf{x}},t; 0,0)]
\right\rangle ,
\end{equation}
where $M_u$ and $M_s$ are Dirac matrices  for light $u$ quark and $s$ quark,
respectively, and the trace runs over the color index.

As explained in Ref.~\refcite{Fu:2011xb},
in principle, bubble contribution\cite{Prelovsek:2005rf}
should be incorporated into the lattice correlator in Eq.~(\ref{CCEQ_kappa}),
\begin{equation}
\label{Ctot_kappa}
C(t)=Ae^{-m_{\kappa}t}+B_{\kappa}(t),
\end{equation}
where, for easier notation, we do not explicitly write down
the contributions from the excited $\kappa$ meson,
and the oscillating terms.

The bubble contribution $B_{\kappa}$ is provided
in the momentum space by Eq.~(15) in Ref.~\refcite{Fu:2011xb}.
The time-Fourier transform of it yields $B_{\kappa}(t)$, namely,
\begin{equation}
B_{\kappa}(t)  =
\frac{\mu^2 }{ 4L^3 }
\bigg\{ f_{B}(t) + f_{V}(t) + f_{A}(t) \bigg\},
\label{eQ:kappaB}
\end{equation}
where $\mu=m_\pi^2/(2m_u)$, and
\begin{eqnarray}
f_{V}(t)  &\equiv& \sum_{\bf k} \left\{
C_{V_{\eta }}^2 \frac{e^{-\left(\sqrt{M_{K_V}^2 + \mathbf{k}^2} +
\sqrt{M_{\eta_V}^2 + \mathbf{k}^2}\right)t} }
{\sqrt{M_{K_V}^2 + \mathbf{k}^2}  \sqrt{M_{\eta_V}^2 + \mathbf{k}^2} }
+C_{V_{\eta'}}^2\frac{e^{-\left(\sqrt{M_{K_V}^2 + \mathbf{k}^2} +
\sqrt{M_{\eta'_V}^2 + \mathbf{k}^2}\right)t} }
{\sqrt{M_{K_V}^2 + \mathbf{k}^2}  \sqrt{M_{\eta'_V}^2 + \mathbf{k}^2} }  \right.\cr
&& \left.
-2\frac{e^{-\left(\sqrt{M_{K_V}^2 + \mathbf{k}^2} +
\sqrt{M_{U_V}^2 + \mathbf{k}^2}\right)t} }
{\sqrt{M_{K_V}^2 + \mathbf{k}^2}  \sqrt{M_{U_V}^2 + \mathbf{k}^2} }
-2\frac{e^{-\left(\sqrt{M_{K_V}^2 + \mathbf{k}^2} +
\sqrt{M_{S_V}^2 + \mathbf{k}^2}\right)t} }
{\sqrt{M_{K_V}^2 + \mathbf{k}^2}  \sqrt{M_{S_V}^2 + \mathbf{k}^2} }
\right\},
\end{eqnarray}

\begin{eqnarray}
\hspace{-0.6cm}
f_{B}(t)  &\equiv& {\sum_{\bf k}} \left\{
\frac{2}{3} \frac{e^{-\left(\sqrt{M_{K_I}^2 + \mathbf{k}^2} +
\sqrt{M_{\eta_I}^2 + \mathbf{k}^2}\right)t} }
{\sqrt{M_{K_I}^2 + \mathbf{k}^2}  \sqrt{M_{\eta_I}^2 + \mathbf{k}^2} }
-2 \frac{e^{-\left(\sqrt{M_{K_I}^2 + \mathbf{k}^2} +
\sqrt{M_{U_I}^2 + \mathbf{k}^2}\right)t} }
{\sqrt{M_{K_I}^2 + \mathbf{k}^2} \sqrt{M_{U_I}^2 + \mathbf{k}^2} }
\right. \cr
&&
\left.
-2 \frac{e^{-\left(\sqrt{M_{K_I}^2 + \mathbf{k}^2} +
\sqrt{M_{S_I}^2 + \mathbf{k}^2}\right)t} }
{\sqrt{M_{K_I}^2 + \mathbf{k}^2}  \sqrt{M_{S_I}^2 + \mathbf{k}^2} }
+\frac{1}{2}\sum_{b=1}^{16}
\frac{e^{-\left(\sqrt{M_{K_b}^2 + \mathbf{k}^2} +
\sqrt{M_{U_b}^2 + \mathbf{k}^2}\right)t} }
{\sqrt{M_{K_b}^2 + \mathbf{k}^2} \sqrt{M_{U_b}^2 + \mathbf{k}^2} }
\right. \cr
&&
\left.
+\frac{1}{4}\sum_{b=1}^{16}
\frac{e^{-\left(\sqrt{M_{K_b}^2 + \mathbf{k}^2} +
\sqrt{M_{s_b}^2 + \mathbf{k}^2}\right)t} }
{\sqrt{M_{K_b}^2 + \mathbf{k}^2}  \sqrt{M_{s_b}^2 + \mathbf{k}^2} }
\right\},
\label{btt:a0}
\end{eqnarray}
where
\begin{equation}
C_{V_\eta}    = 2\frac{Z_V - 5\delta_V/4}{Z_V}, \qquad
C_{V_{\eta'}} = 2\frac{Z_V + 5\delta_V/4}{Z_V} ,
\label{FZW_03:appC}
\end{equation}
for $f_{A}(t)$, we just require
$V \to A$ in $f_{V}(t)$.
$M_{\eta V}$, $M_{\eta A}$, $M_{\eta' V}$, $M_{\eta' A}$, $Z_{V}$ ,$Z_{A}$ are
given in Eq.(~\ref{eq_etaAV}).

\section{Simulations and results}
\label{sec_results}
We employ the MILC $2+1$ flavors gauge configurations with 
the Asqtad-improved staggered sea quarks.
See more descriptions in Refs.~\refcite{Bazavov:2009bb,Bernard:2010fr,Aubin:2004wf}.
We processed the $\kappa$ propagators on the $0.09$~fm MILC fine lattice ensemble
of $500$ $28^3 \times 96$ gauge configurations
with bare quark masses $am_{ud}'/am_s' = 0.0062/0.031$,
bare gauge coupling $10/g^2 = 7.09$
and the inverse lattice spacing $a^{-1}=2.349_{-23}^{+61}$~GeV.
The dynamical strange quark mass is near to its physical mass,\cite{Bazavov:2009bb,Bernard:2010fr}
and the light $u/d$ quark masses are degenerate.
In Table~\ref{tab_L0}, we tabulate the pseudoscalar masses
needed in this work except the masses $M_{\eta_A}$, $M_{\eta^\prime_A}$,
$M_{\eta_V}$ and $M_{\eta^\prime_V}$,
which are changeable with the fit parameters $\delta_A$ and $\delta_V$.

We make use of the conjugate gradient method to get the necessary
matrix element of the fermion matrix $M_u^{-1}$ and $M_s^{-1}$.
Then we make use of Eq.~(\ref{CCEQ_kappa}) to
measure the point-to-point $\kappa$ correlator.
To enhance the statistics, we put the source
on all the time slices,
namely, we carry out $T=96$ matrix inversions per gauge configuration
and average these propagators after measurement.
This rather large of matrix inversions  
enables us to compute the $\kappa$ correlators with acceptable accuracy,
which is crucial to to our ultimate results.

Since the $\kappa$ meson comprises a strange $s$ quark and a light $u$ quark,
the $u$ quark is usually handled as a valence approximation quark,
on the other hand the valence strange quark mass is
set to be its physical mass,\cite{Aubin:2004fs}
which was robustly measured
by the MILC Collaboration.\cite{Bazavov:2009bb,Bernard:2010fr}

Using the same configurations,
the $\kappa$ correlators are evaluated for five $u$ valence quarks,
to be specific, we select $am_x = 0.0062$, $0.0093$, $0.0124$, $0.0155$ and $0.0186$,
where $m_x$ is the valence up quark mass.
To secure the physical $\kappa$ mass, we carry out
the extrapolation to the physical limit (physical $\pi$ mass quoted from PDG).
The propagators of the $\pi, K$ and artificial $s\bar{s}$ meson
are calculated with the same gauge configurations
to compute the pseudoscalar masses in Table~\ref{tab_L0}.

For staggered quarks the meson propagators hold generic functional form,
\begin{equation}
\label{sfits:ch7}
{\cal C}(t) =
\sum_i A_i e^{-m_i t} + \sum_i A_i^{\prime}(-1)^t e^{-m_i^{\prime} t}  +(t \rightarrow N_t-t),
\end{equation}
where the oscillating terms stand for a particle with contrary parity.
In practice, for the $\kappa$ meson propagator,
only one mass with each parity is considered to get acceptable fits.\cite{Bernard:2007qf}
In the presence of the bubble contribution,
all five $\kappa$ propagators are fit with the below physical model
\begin{equation}
C_{\kappa}(t) = C_{\kappa}^{\rm meson} (t)  + B_{\kappa}(t),
\label{eq:fitfcn}
\end{equation}
here
\begin{equation}
  C_{\kappa}^{ \rm meson} (t) = b_{\kappa}e^{-m_{\kappa}t} +
  b_{K_A}(-1)^t e^{-M_{K_A}t} + (t \rightarrow N_t-t), \nonumber
\end{equation}
where $b_{K_A}$ and $b_{\kappa}$ are two overlap amplitudes,
and the bubble contribution $B_{\kappa}(t) $
is provided in Eq.~(\ref{eQ:kappaB}).

The above fitting model consists of a $\kappa$ pole,
along with the corresponding opposite-parity state $K_A$
and the bubble term.\cite{Fu:2011xb}
For a given $m_x$, there exist four fit parameters
(i.e., $M_{\kappa}$, $M_{K_A}$, $b_{K_A}$,
and $b_{\kappa}$) for each $\kappa$ pole,
but the opposite parity masses $K_A$ were closely restricted
by priors to be the same calculated masses which are substituted in the bubble term.
The bubble contribution $B_{\kappa}(t)$ was
parameterized by three low-energy coupling constants $\mu$, $\delta_A $,
and $\delta_V $, which were allowed to vary to get the best fit.
The taste multiplet masses in the bubble terms were
fixed to be the values as tabulated in Table~\ref{tab_L0}.
The summation over intermediate momenta was cut off if
either the total energy of the two-body state surpassed $2.0/a$ or
any momentum  constituent surpassed  $\pi/(4a)$.
Such a cutoff is turned out to yield a 
satisfactory result for $t \ge 8$.

In practice, the $\kappa$ masses are extracted
from the effective mass plots, and they were selected
by the comprehensive consideration of a ``plateau'' in the effective mass plots
as a function of the minimum distance $D_{\rm min}$,
a good confidence level (i.e.,$\chi^2$) for the fit
and $D_{\rm min}$ large enough to suppress the
excited states.\cite{Fu:2011xb}
We note that the effective $\kappa$ mass often suffers from large statistical errors,
particularly in the region of large $D_{\rm min}$.
To avert possible large errors stemming from the data at large $D_{\rm min}$,
we fit the $\kappa$ propagators only
in the time range $12\le D_{\rm min} \le 16$,
where the effective masses are reaching a plateau with relatively small errors.
In reality, five $\kappa$ correlators were fit with $D_{\rm min}=15a$.
At this distance,  the systematic effect due to the excited states
can be reasonably neglected.\cite{Fu:2011xb}

The fitted $\kappa$ masses are listed in Table~\ref{fitted_table}.
Column $2$ gives the $\kappa$ masses in lattice units,
and Column $4$ shows the fit range.
As a consistency check, we present the fitted $K_A$  masses
in Column $3$ as well.
It is important to notice that the fitted $K_A$ masses
are in well agreement with our computed masses
listed in Table~\ref{tab_L0} within rather small errors.
Column $5$ gives fit quality $\chi^2/{\rm d.o.f}$.

\begin{table}[h]
\tbl{\label{fitted_table}
Summaries of the fitted $\kappa$ masses.
The Column $2$ gives the fitted $\kappa$ masses in lattice units.
the fitted $K_A$ masses are presented in Column $3$.
}
{\begin{tabular}{ccccc}
\toprule
$am_x$  & $am_\kappa$  & $aM_{K_A}$ & {\rm Range} & $\chi^2/{\rm dof}$  \\
\colrule
$0.0062$ & $0.439(7)$ & $0.2441(27)$ & $15-25$ & $2.5/4$ \\
$0.0093$ & $0.479(8)$ & $0.2547(25)$ & $15-25$ & $3.7/4$ \\
$0.0124$ & $0.508(9)$ & $0.2650(22)$ & $15-25$ & $4.7/4$ \\
$0.0155$ & $0.526(8)$ & $0.2749(21)$ & $15-25$ & $5.6/4$ \\
$0.0186$ & $0.539(8)$ & $0.2846(20)$ & $15-25$ & $6.3/4$ \\
\botrule
\end{tabular}}
\end{table}

To get the physical $\kappa$ mass, we carry out the chiral extrapolation
of the $\kappa$ mass $m_\kappa$ to the physical point
through the popular three fit parameters together with
the chiral logarithms.\cite{Nebreda:2010wv}
The generic functional form of the pion mass dependence of $m_\kappa$
is expressed as
\begin{equation}
m_\kappa = c_0 + c_2 m_\pi^2 + c_3 m_\pi^3 + c_4 m_\pi^4\ln(m_\pi^2),
\end{equation}
where $c_0, c_2, c_3$ and $c_4$ are four fitting parameters,
and the last term is the chiral logarithms.

\begin{figure}[h!]
\begin{center}
\includegraphics[width=8.0cm]{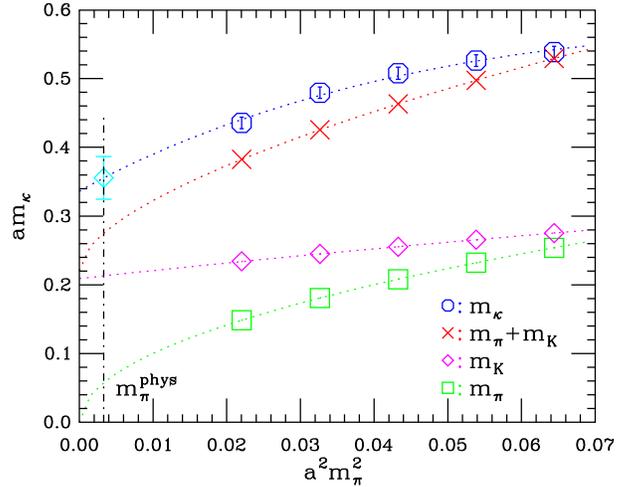}
\end{center}
\caption{\label{fig:kappa_limit}
Features of $m_\kappa, m_K, m_\pi$ and $m_\pi\hspace{-0.05cm}+\hspace{-0.05cm}m_K$
in lattice units as a function of pion mass.
The physical $\kappa$ mass is obtained at the physical pion mass $m_\pi$.
}
\end{figure}

We quote the physical pion mass from PDG\cite{Beringer:1900zz} as the physical limit.
In Fig.~\ref{fig:kappa_limit},
we demonstrate  how physical kappa mass $m_\kappa$ is secured,
which yields $\chi^2/{\rm dof}=0.97/1$.
The blue dashed line in Fig.~\ref{fig:kappa_limit}
is the chiral extrapolation of $\kappa$ mass
to physical point.
The chirally extrapolated $\kappa$ mass $m_\kappa=(835\pm93)$~MeV,
which is consistent with our previous study
on a MILC ``coarse'' lattice ensemble~\cite{Fu:2011xb}, and
is in agreement with the BES experimental results.\cite{Ablikim:2010ab,Ablikim:2010kd}
The cyan diamond in Fig.~\ref{fig:kappa_limit} displays our extracted  physical $\kappa$ mass.
In this same figure, we show kaon masses $m_K$, pion masses $m_\pi$,
and $m_{\pi}+m_K$ in lattice units as a function of  $m_\pi$ as well.

To understand the effects of the bubble contribution,
in this work we also fitted our measured $\kappa$ correlators without bubble terms.
The negative parity masses $K_A$ were tightly restricted by priors
to be the derived mass used in the bubble term.
The fitted results are tabulated in Table~\ref{fitted_table_no_bubble}.
From Tables~\ref{fitted_table} and~\ref{fitted_table_no_bubble},
we can clearly see that the bubble terms contribute
about $1\%$ differences for the $\kappa$ masses.
In our previous work on a MILC ``coarse'' lattice ensemble ($a=0.12$~fm),\cite{Fu:2011xb}
the bubble terms contribute about $2\%-5\%$ difference for the kappa mass,
and we refitted the lattice data in our previous study on a MILC ``medium-coarse''
lattice ensemble ($a=0.15$~fm) with the inclusion of bubble contribution,
and found that the differences are as large as about $3\%-8\%$.\cite{Fu:2011zz}
These results are what we expected, since the bubble contribution is
a kind of the lattice artifacts caused by the fourth-root approximation,\cite{Fu:2011gw,Bernard:2007qf}
and the artifacts contain the thresholds at unphysical energies and
the thresholds with negative weights.\cite{Fu:2011xb}
The rS$\chi$PT\ forecasts  that these lattice artifacts
disappear in the continuum limit, only keeping physical two-body thresholds.
Therefore, it is a natural aftermath that the bubble contribution
becomes less important as the lattice spacing $a$
used in the lattice ensembles is smaller.
Although we have minimized the lattice artifacts by
using a MILC fine $a \approx 0.09$~fm lattice ensemble,
an empirical investigation of these effects is still highly desired.
We will apply all the possible computer resources to investigate
whether this expectation is gotten rid of in lattice simulations
at the much smaller lattice spacing, e.g., $a \approx 0.06,0.045$~fm
in the future.

\begin{table}[ht]
\tbl{\label{fitted_table_no_bubble}
Summaries of the $\kappa$ masses fitted without bubble contributions.
Column $2$ presents the fitted $\kappa$ masses in lattice units.
Column $3$ gives the fitted $K_A$ masses.
}
{\begin{tabular}{ccccc}
\toprule
$am_x$  & $am_\kappa$  & $aM_{K_A}$ & {\rm Range} & $\chi^2/{\rm dof}$  \\
\colrule
$0.0062$ & $0.440(14)$ & $0.2440(27)$ & $15-25$ & $2.7/7$ \\
$0.0093$ & $0.480(13)$ & $0.2547(25)$ & $15-25$ & $3.8/7$ \\
$0.0124$ & $0.505(13)$ & $0.2650(23)$ & $15-25$ & $4.8/7$ \\
$0.0155$ & $0.522(12)$ & $0.2749(21)$ & $15-25$ & $5.6/7$ \\
$0.0186$ & $0.534(10)$ & $0.2846(20)$ & $15-25$ & $6.3/7$ \\
\botrule
\end{tabular}}
\end{table}

\section{Summary}
\label{sec_conclude}
In Ref.~\refcite{Fu:2011xb}, we derived bubble contribution
to $\kappa$ correlator in the lowest order S$\chi$PT.
We used this physical model to fit the $\kappa$ propagators
for a MILC fine ($a\approx0.09$~fm) lattice ensemble
with the $2+1$ flavors of the Asqtad-improved staggered sea quarks.
We familiarly handled the light $u$ quark as a valence approximation quark,
whereas the strange valence quark mass is set at its physical mass,
and chirally extrapolated the $\kappa$ mass to the physical point.
We achieved the $\kappa$ mass with $835\pm93$~MeV,
which is consistent with our previous work.\cite{Fu:2011zz,Fu:2011xb}
Additionally, our simulation results demonstrated
that the bubble contribution is negligible in the accuracy of
statistics for the MILC fine lattice ensemble used in this work.


\section*{Acknowledgments}
This work is in part supported by Fundamental Research Funds for
the Central Universities (2010SCU23002).
We would thank the MILC Collaboration for using
the Asqtad lattice ensemble and MILC code.
We are grateful to Hou Qing for his support.
The computations for this work were carried out at AMAX, CENTOS, HP workstations
in the Radiation Physics Group of the Institute of
Nuclear Science and Technology, Sichuan University.


\end{document}